\documentclass[11pt,a4paper,final]{article}
\usepackage{epsfig}
\usepackage{amsmath}
\usepackage{graphics}

\def\nue{\nu_{e}}
\def\num{\nu_{\mu}}
\def\nut{\nu_{\tau}}
\def\nus{\nu_{sterile}}
\def\ee{e^{+}e^{-}}
\def\ll{\langle}
\def\rr{\rangle}
\def\aprle{\buildrel < \over {_{\sim}}}

\def\ra{\rightarrow}

\def\be{\begin{equation}}
\def\ee{\end{equation}} 
\def\bea{\begin{eqnarray}}
\def\eea{\end{eqnarray}}

\relax
\title{Search for a Lorentz invariance violation contribution in atmospheric neutrino oscillations using MACRO data.}

\begin{document}

\date{}

\maketitle

%%%%%%%%%%%%%%%%%%%%%%%%%%%%%%%%%%%%%%
\vspace{-7.00cm}
\hspace{9cm}
DFUB 02/2005

\hspace{9cm}
Bologna, 07/03/05
\vspace{5.00cm}
%%%%%%%%%%%%%%%%%%%%%%%%%%%%%%%%%%%%%%

\author{
G.~Battistoni$^{a}$, \and
Y.~Becherini$^{b}$, \and
S.~Cecchini$^{b}$, \and
M.~Cozzi$^{b}$, \and
H.~Dekhissi$^{b,c}$, \and
L.~S.~Esposito$^{b}$, \and
G.~Giacomelli$^{b}$, \and
M.~Giorgini$^{b}$, \and
G.~Mandrioli$^{b}$, \and
S.~Manzoor$^{b,d}$, \and
A.~Margiotta$^{b}$, \and
L.~Patrizii$^{b}$, \and
V.~Popa$^{b,e}$, \and
M.~Sioli$^{b}$, \and
G.~Sirri$^{b}$, \and
M.~Spurio$^{b}$, \and
V.~Togo$^{b}$.
}

\begin{center}
$^{a}$INFN Sezione di Milano, Via Celoria 16, I-20133 Milano, Italy \\
$^{b}$ Dipartimento di Fisica dell'Universit\`a di Bologna and \\
INFN Sezione di Bologna, Viale Berti Pichat 6/2, I-40127 Bologna, Italy\\
$^{c}$Also Facult\`e des Sciences, Universit\`e Mohamed Ier, Oujda, Morocco\\
$^{d}$Also RPD, PINSTECH, P.O. Nilore, Islamabad, Pakistan\\
$^{e}$Also Institute for Space Sciences, R77125 Bucharest, Romania\\
\end{center}

\vspace{0.5cm}

\abstract{
\noindent Neutrino-induced upward-going muons in MACRO have been analysed 
in terms of relativity principles violating effects, keeping standard 
mass-induced atmospheric neutrino oscillations as the dominant source of 
$\num \ra \nut$ transitions. The data disfavor these exotic
possibilities even at a sub-dominant level, and stringent 90\% C.L. 
limits are placed on the Lorentz invariance violation parameter 
$|\Delta v| < 6 \times 10^{-24}$ at $\sin 2{\theta}_v$ = 0 and
$|\Delta v| < 2.5 \div 5 \times 10^{-26}$ at $\sin 2{\theta}_v$ = $\pm$1.
These limits can also be re-interpreted as upper bounds on the
parameters describing violation of the Equivalence Principle.
}

\vspace{1cm}

Neutrino mass-induced oscillations are the best explanation of the atmospheric neutrino 
problem \cite{macro-98, macro-last, sk-dip, soudan2}. Two flavor $\num \ra \nut$
oscillations are strongly favored over a wide range of alternative solutions such as
$\num \ra \nus$ oscillations \cite{macro-sterile, sk-sterile}, 
$\num \ra \nue$ oscillations \cite{sk-dip, soudan2}
or other exotic possibilities \cite{sk-habig, vlad-oujda}.

In this letter, we assume standard mass-induced neutrino oscillations as
the leading mechanism for flavor transitions and we treat Lorentz invariance
flavor transitions as a sub-dominant effect \cite{fogli}.
In the literature, neutrino oscillations induced by Violation of
(CPT-conserving) Lorentz Invariance (VLI) and 
Violation of the Equivalence Principle (VEP) are
described within the same formalism. 
In the following we will mention only VLI for simplicity.

In this scenario, neutrinos can be described in terms of three distinct
bases: flavor eigenstates, mass eigenstates and velocity eigenstates,
the latter being characterized by different Maximum Attainable Velocities (MAVs) 
in the limit of infinite momentum.

\vspace{0.5cm}

Both mass-induced oscillations and VLI transitions are treated in the
two-family approximation and we assume that mass and velocity mixings 
occur inside the same families (e.g. $| \nu_2 \rangle$ and $| \nu_3 \rangle$).

\vspace{0.5cm}

The usual interpretation of the atmospheric neutrino oscillations is $\num \ra \nut$ 
induced by the mixing of the two {\it mass} eigenstates  $| \nu^m_2 \rangle$ and $| \nu^m_3 \rangle$,
and two weak eigenstates  $| \num \rangle$ and $| \nut \rangle$, i.e.

\begin{equation}
\begin{array}{ll}
  | \nu_\mu  \rangle =  \ \ |\nu^m_2 \rangle \cos \theta^m_{23} + | \nu^m_3 \rangle \sin \theta^m_{23} \\[6pt]
  | \nu_\tau \rangle = -|\nu^m_2 \rangle \sin \theta^m_{23} + | \nu^m_3 \rangle \cos \theta^m_{23}
\end{array}
\end{equation}

\noindent where $\theta^m_{23}$ ($\equiv \theta_m$) is the flavor-mass mixing angle. The survival
probability of muon neutrinos at a distance $L$ from production is

\begin{equation}
        P(\num \ra \num) = 1 - \sin^2 2\theta_m \sin^2 (1.27 \Delta m^2 L/E_{\nu})
\label{eq:prob1}
\end{equation}

\noindent where $\Delta m^2 = ( m^2_{\nu^m_3} - m^2_{\nu^m_2} )$
is expressed in eV$^2$, $L$ in km and the neutrino energy $E_{\nu}$ in GeV.
Notice the dependence on $L/E_{\nu}$ in the argument of the second sin$^2$ term.

In the VLI case, the two flavor eigenstates $| \nu_\mu \rangle$, $| \nu_\tau \rangle$
and the two velocity eigenstates $| \nu^v_2 \rangle$, $| \nu^v_3 \rangle$
are connected through the mixing angle $\theta^v_{23}$ ($\equiv \theta_v$)
in analogy with mass-induced oscillations:

\begin{equation}
\begin{array}{ll}
  | \num  \rangle =  \ \ |\nu^v_2 \rangle \cos \theta^v_{23} + | \nu^v_3 \rangle \sin \theta^v_{23} \\[6pt]
  | \nut \rangle = -|\nu^v_2 \rangle \sin \theta^v_{23} + | \nu^v_3 \rangle \cos \theta^v_{23}
\end{array}
\end{equation}

\noindent In this case, the $\num$ survival probability is 

\begin{equation}
        P(\num \ra \num) = 1 - \sin^2 2\theta_v \sin^2 (2.54 \! \cdot \!10^{18} \Delta v \ L E_{\nu})
\label{eq:prob1b}
\end{equation}

\noindent where $\Delta v = ( v_{\nu^v_3} - v_{\nu^v_2} )$ is the neutrino
MAV difference in units of $c$.
Notice that neutrino flavor oscillations induced by VLI are characterized by an 
$L E_{\nu}$ dependence of the oscillation probability (Eq. \ref{eq:prob1b}),
to be compared with the $L/E_{\nu}$ behavior of mass-induced oscillations (Eq. \ref{eq:prob1}).

When both mass-induced transitions and VLI-induced transitions are
considered simultaneously, the muon neutrino survival probability can be 
expressed as \cite{fogli,glashow9799,glashow04}

\begin{equation}
        P(\num \ra \num) = 1 - \sin^2 2\Theta \sin^2 \Omega
\label{eq:prob2}
\end{equation}

\noindent where the global mixing angle $\Theta$ and the term $\Omega$ are
given by:

\begin{equation}
\begin{array}{ll}
2 \Theta = \mathrm{atan} (a_1/a_2) \\[6pt]
\Omega = \sqrt{(a_1^2+a_2^2)}
\label{eq:mix}
\end{array}
\end{equation}
with
\begin{equation}
\begin{array}{ll}
  a_1 = 1.27 \left| \Delta m^2 \sin 2{\theta}_m L/E_{\nu} + 
  2\! \cdot \!10^{18} \Delta v \sin 2{\theta}_v \ L E_{\nu} \ e^{i\eta} \right| \\[6pt]
  a_2= 1.27 \left( \Delta m^2 \cos 2{\theta}_m L/E_{\nu} + 
  2\! \cdot \!10^{18} \Delta v \cos 2{\theta}_v \ L E_{\nu} \right)
\end{array}
\end{equation}

\noindent Here $\Delta m^2$, $L$ and $E_{\nu}$ are expressed, 
as in Eq. \ref{eq:prob1} and Eq. \ref{eq:prob1b},
in eV$^2$, km and GeV, respectively. The additional factor $e^{i\eta}$
connects the mass and velocity eigenstates, and for the moment it is assumed to
be real ($\eta$ = 0 or $\pi$). Note that formulae \ref{eq:prob1} 
and \ref{eq:prob2} do not depend on the sign of the mixing angle and/or
on the sign of the $\Delta v$ and $\Delta m^2$ parameters; this is not so 
in the case of mixed oscillations, where the relative sign between
the mass-induced and VLI-induced oscillation terms is important.
The whole domain of variability of the parameters can be accessed with the requirements 
$\Delta m^{2} \ge 0$, $0 \le \theta_m \le \pi/2$, $\Delta v \ge 0$
and $-\pi/4 \le \theta_v \le \pi/4$.

The same formalism also applies to violation of the equivalence principle,
after substituting $\Delta v/2$ with the adimensional product $|\phi| \Delta \gamma$; 
$\Delta \gamma$ is the difference of the coupling constants for neutrinos 
of different types to the gravitational potential $\phi$ \cite{gasperini}.

\begin{figure}[p]
\begin{center}
\includegraphics[scale=0.8,bb=70 -20 500 500]{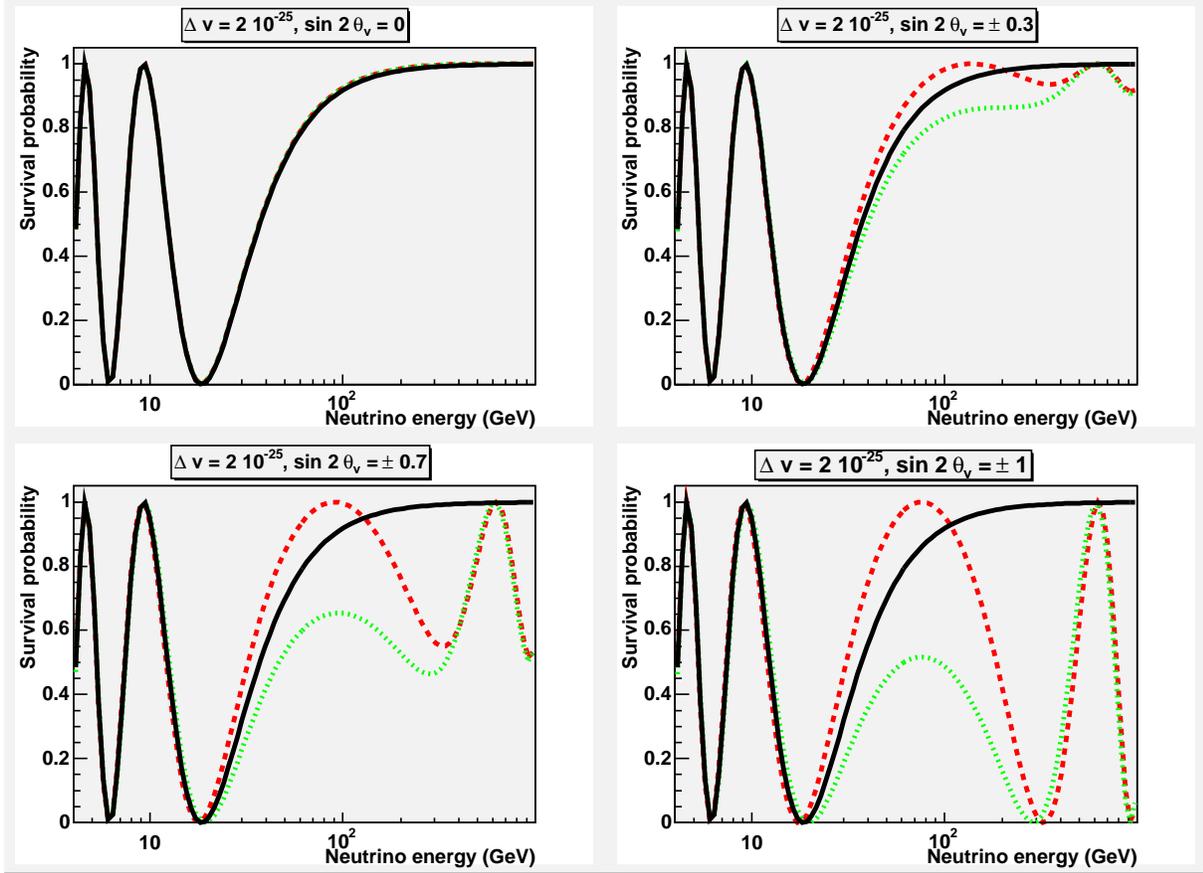}
\caption{
  Energy dependence of the $\num \ra \num$ survival probability for 
  mass-induced oscillations alone (continuous line) 
  and mass-induced + VLI oscillations for $\Delta v = 2 \cdot 10^{-25}$ 
  and sin $2\theta_v$ = 0, $\pm$0.3, $\pm$0.7 and $\pm$1 
  (dashed lines for positive values, dotted lines for negative values).
  The neutrino pathlength was fixed at $L$ = 10000
  km and we assumed $\Delta m^2$ = 0.0023 eV$^2$, $\theta_m$ = $\pi/4$.
}
\label{f:ftest}
\end{center}
\end{figure}

As shown in \cite{glashow9799}, and more recently in \cite{glashow04},
the most sensitive tests of VLI can be made by analysing the high energy tail 
of atmospheric neutrinos at large pathlength values. As an example,
Fig. \ref{f:ftest} shows the energy dependence of the $\num \ra \num$ survival
probability as a function of the neutrino energy, for neutrino mass-induced 
oscillations alone and for both mass and VLI-induced oscillations for
$\Delta v = 2 \times 10^{-25}$ and different values of sin 2$\theta_v$ parameter.
Note the large sensitivity for large neutrino energies and large mixing angles.
Given the very small neutrino mass ($m_{\nu} \aprle 1$ eV), neutrinos with
energies larger than 100 GeV are extremely relativistic, with Lorentz
$\gamma$ factors larger than $10^{11}$.

\vspace{0.5cm}

MACRO \cite{macrodet} was a multipurpose large area detector 
($\sim$10000 m$^2$ sr acceptance for an isotropic flux)
located in the Gran Sasso underground Lab,
shielded by a minimum rock overburden of 3150 hg/cm$^2$.
The detector had global dimensions of 76.6 $\times$ 12 $\times$ 9.3 m$^3$
and used limited streamer tubes and scintillation counters to detect
muons. $\num$'s were detected via charged current interactions 
$\num + N \ra \mu + X$; upgoing muons were identified with the
streamer tube system (for tracking) and the scintillator system (for
time-of-flight measurement).
Early results concerning atmospheric neutrinos were published in 
\cite{macro-95} and in \cite{macro-98} for the upthroughgoing muon sample
and in \cite{macro-le} for the low energy semi-contained and
upgoing-stopping muon events.
Matter effects in the $\num \ra \nus$ channel were presented
in \cite{macro-sterile} and a global analysis of all MACRO neutrino 
data in \cite{macro-last}.

\vspace{0.5cm}

In order to analyse the MACRO data in terms of VLI, we used a subsample 
of 300 upthroughgoing muons whose energies were estimated via 
Multiple Coulomb Scattering in the 7 horizontal rock absorbers in the lower
apparatus \cite{mcs1,mcs2}. The energy estimate was obtained
using the streamer tubes in drift mode, which allowed to
considerably improve the spatial resolution of the detector ($\sim$ 3 mm). 
The overall neutrino energy resolution was of the order of 100\%, 
mainly dominated by muon energy losses in the rock below the detector
(note that $\ll E_{\mu} \rr \simeq 0.4 \ \ll E_{\nu} \rr$).
Upgoing muon neutrinos of this sample have large zenith angles 
($> 120^{\circ}$) and the median value of neutrino pathlengths 
is slightly larger than 10000 km.

Following the analysis in Ref. \cite{mcs2}, we selected a low and a high energy sample
by requiring that the reconstructed neutrino energy $E^{rec}_\nu$ should be
$E^{rec}_\nu <$ 30 GeV and $E^{rec}_\nu >$ 130 GeV.
The number of events surviving these cuts is $N_{low}$ = 49 and $N_{high}$ =
58, respectively; their median energies, estimated via Monte Carlo,
are 13 GeV and 204 GeV (assuming mass-induced oscillations).

The analysis then proceeds by fixing the neutrino mass oscillation
parameters at the values obtained with the global analysis of all MACRO neutrino data
\cite{macro-last}:
$\overline{\Delta m^2}$ = 0.0023 eV$^2$, sin$^2$2$\overline{\theta}_m$ = 1.
Then, we scanned the plane of the two free parameters 
($\Delta v$, $\theta_v$) using the function

\begin{equation}
  \chi^2 = \sum_{i=low}^{high} \left(\frac{N_i-\alpha 
    N_i^{MC}(\Delta v, \theta_v; \overline{\Delta m^2}, \overline{\theta}_m)}{\sigma_i} \right)^2
\end{equation}

\noindent where $N_i^{MC}$ is the number of events predicted by Monte Carlo,
$\alpha$ is a constant which normalizes the number of Monte Carlo events 
to the number of observed events and $\sigma_i$ is the overall error
comprehensive of statistical and systematic uncertainties.

We used the Monte Carlo simulation described in \cite{mcs2} with different
neutrino fluxes in input \cite{bartol, fluka01, honda01, newhonda}.
The largest relative difference of the extreme values of the MC expected 
ratio $N_{low}/N_{high}$ is 13\%.
However, in the evaluation of the systematic error, the main sources
of uncertainties for this ratio
(namely the primary cosmic ray spectral index and neutrino cross sections)
have been separately estimated and their effects added in quadrature 
(see \cite{mcs2} for details):
in this work, we use a conservative 16\% theoretical systematic error 
on the ratio $N_{low}/N_{high}$.
The experimental systematic error on the ratio was estimated to be 6\%.
In the following, we show the results obtained with the
computation in \cite{newhonda}.

\begin{figure}[p]
\begin{center}
\includegraphics[scale=0.8,bb=80 80 500 500]{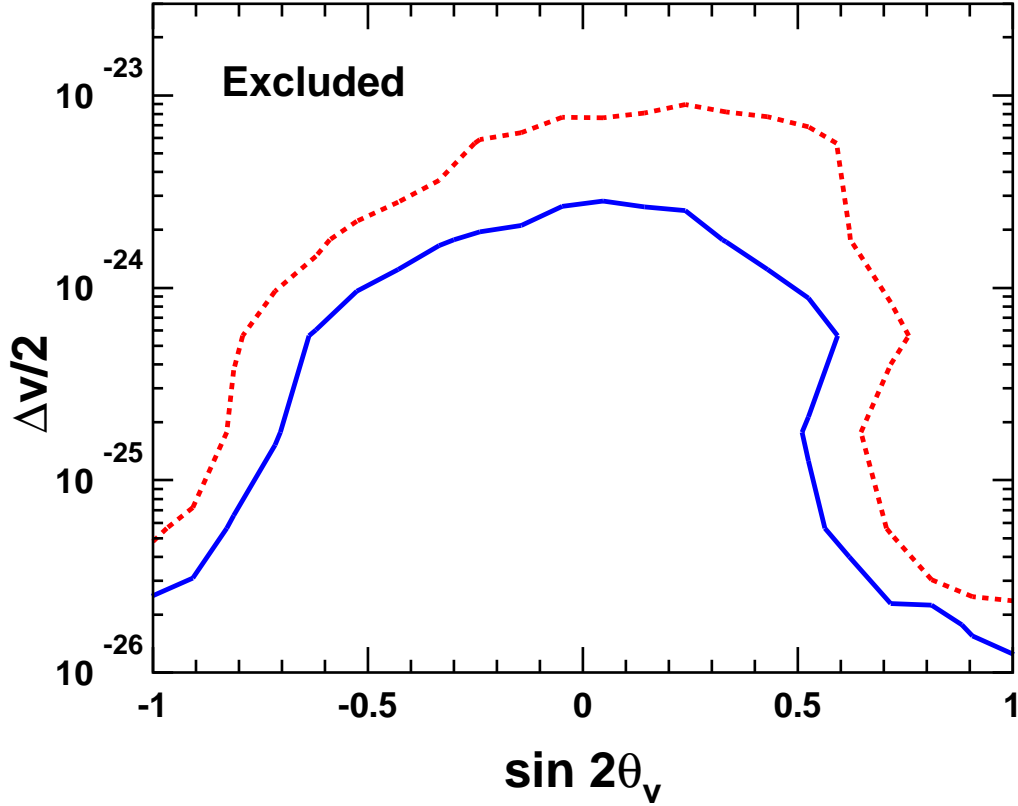}
\caption{
  90\% C.L. upper limits on the Lorentz invariance violation
  parameter $\Delta v$ versus sin $2\theta_v$.
  Standard mass induced oscillations are assumed in
  the two-flavor $\num \ra \nut$ approximation, 
  with $\Delta m^2$ = 0.0023 eV$^2$ and $\theta_m$ = $\pi/4$.
  The dashed line shows the limit obtained with the same selection criteria of
  Ref. \cite{mcs2} to define the low and high energy samples; the
  continuous line is the final result obtained with the selection criteria 
  optimized for the present analysis (see text).
}
\label{f:super}
\end{center}
\end{figure}

The inclusion of the VLI effect does not improve the $\chi^2$ in any point
of the ($\Delta v$, $\theta_v$) plane, compared to mass-induced oscillations
stand-alone, and proper upper limits on VLI parameters were obtained. 
The 90\% C.L. limits on $\Delta v$ and
$\theta_v$, computed with the Feldman and Cousins prescription \cite{felcou}, 
are shown by the dashed line in Fig. \ref{f:super}.

The energy cuts described above (the same used in Ref. \cite{mcs2}), 
were optimized for mass-induced neutrino oscillations. In order to
maximize the sensitivity of the analysis for VLI induced oscillations,
we performed a blind analysis, based only on Monte Carlo events, to 
determine the energy cuts which yield the best performances. The results of this study 
suggest the cuts $E^{rec}_\nu <$ 28 GeV and $E^{rec}_\nu >$ 142 GeV;
with these cuts the number of events in the real data are
$N^{\prime}_{low}$ = 44 events and $N^{\prime}_{high}$ = 35 events.
%and to median energies of 34 GeV and 216 GeV, respectively (assuming mass-induced oscillations).
The limits obtained with this selection are shown in
Fig. \ref{f:super} by the continuous line.
As expected, the limits are now more stringent than for the previous choice.

In order to understand the dependence of this result with respect to the
choice of the $\overline{\Delta m^2}$ parameter, we varied the
$\overline{\Delta m^2}$ values around the best-fit point. We found that a variation
of $\overline{\Delta m^2}$ of $\pm$30\% moves up/down the upper limit of VLI
parameters by at most a factor 2.

Finally, we computed the limit on $\Delta v$ marginalized with respect to 
all the other parameters left free to variate inside the intervals: 
$\Delta m^2$ = $\overline{\Delta m^2} \pm$30\%,
$\theta_m$ = $\overline{\theta}_m$ $\pm$20\%, $-\pi/4 \le \theta_v \le \pi/4$
and any value of the phase $\eta$. We obtained
the 90\% C.L. upper limit $|\Delta v| < 3 \times 10^{-25}$.

\vspace{0.5cm}

An independent and complementary analysis was performed on a sample of 
events with a reconstructed neutrino energy 25 GeV $< E^{rec}_\nu <$ 75 GeV.
The number of events satisfying this condition is 106. A negative 
log-likelihood function was built event by event and then fitted to the
data. We allowed mass-induced oscillation parameters to vary inside the 
MACRO 90\% C.L. region and we left VLI parameters free in the whole 
($\Delta v$, $\theta_v$) plane.
The upper limit on the $\Delta v$ parameter resulting from this analysis
is slowly varying with $\Delta m^2$ and is of the order of $\approx 10^{-25}$.

\vspace{0.5cm}

In conclusion: we have searched for ``exotic'' contributions to standard mass-induced 
atmospheric neutrino oscillations arising from a possible violation of Lorentz
invariance. We used a subsample of MACRO upthroughgoing muon events for
which an energy measurement was made via multiple Coulomb scattering.
The inclusion of VLI effects does not improve the fit to
the data, and we conclude that these effects are disfavored even at
the sub-dominant level.

The 90\% C.L. limits of VLI parameters are 
$|\Delta v| < 6 \times 10^{-24}$ at $\sin 2{\theta}_v$ = 0 and
$|\Delta v| < 2.5 \div 5 \times 10^{-26}$ at $\sin 2{\theta}_v$ = $\pm$1,
see Fig. \ref{f:super}. 
In terms of the parameter $\Delta v$ alone (marginalization with respect 
to all the other parameters), the VLI parameter bound is (at 90\% C.L.) 
$|\Delta v| < 3 \times 10^{-25}$.

These results may be reinterpreted in terms of 90\% C.L. limits of 
parameters connected with violation of the equivalence principle, 
giving the limit $|\phi \Delta \gamma| < 1.5 \times 10^{-25}$.

These limits are comparable or better to those estimated using K2K and 
Super-Kamiokande data \cite{fogli}.

\vspace{0.5cm}

We acknowledge the cooperation of the members of the MACRO Collaboration.
We thank several colleagues for discussions and advise, in particular
B.~Barish, P.~Bernardini, A.~De~Rujula, G.~L.~Fogli, S.~L.~Glashow,
P.~Lipari and F.~Ronga.

%%%%%%%%%%%%%%%%%%%%%%%%%%%%%%%%%%%%%%%%%%%%%%%%%%%%%%%%%%%%%%%%%%%%%%%%%%%%
%%%%%%%%%%%%%%%%%%%%%%%%%%%%%%%%%%%%%%%%%%%%%%%%%%%%%%%%%%%%%%%%%%%%%%%%%%%%

%%%%%%%%%%%%%%%%%%%%%%%%%%%%%%%%%%%%%%%%%%%%%%%%%%%%%%%%%%%%%%%%%%%%%%%%%%%%
%%%%%%%%%%%%%%%%%%%%%%%%%%%%%%%%%%%%%%%%%%%%%%%%%%%%%%%%%%%%%%%%%%%%%%%%%%%%

\end{document}